\documentclass[10pt,DIV12,abstracton,notitlepage]{scrartcl}
\usepackage{mathptmx}
\usepackage{graphicx}
\usepackage{color}
\usepackage[dvips,bookmarksopen,bookmarksnumbered,breaklinks]{hyperref}
\usepackage{breakurl}
\usepackage[final]{listings}
\newcommand\BibTeX{{\rmfamily B\kern-.05em \textsc{i\kern-.025em b}\kern-.08em
T\kern-.1667em\lower.7ex\hbox{E}\kern-.125emX}}
\newcommand{\lperfctr}{\texttt{likwid-perfctr}}
\newcommand{\lbench}{\texttt{likwid-bench}}

\newcommand{\bq}{\begin{equation}}
\newcommand{\eq}{\end{equation}}

\newcommand{\flops}{\mbox{flops}}

\newcommand{\GBS}{\mbox{GB/s}}

\newcommand{\GFS}{\mbox{GF/s}}

\newcommand{\MFS}{\mbox{MF/s}}
\newcommand{\MPS}{\mbox{MP/s}}

\newcommand{\MLUPS}{\mbox{MLUP/s}}

\newcommand{\GHZ}{\mbox{GHz}}
\newcommand{\W}{\mbox{W}}

\newcommand{\BF}{\mbox{B/F}}
\newcommand{\bytes}{\mbox{bytes}}
\newcommand{\byte}{\mbox{byte}}
\newcommand{\bit}{\mbox{bit}}
\newcommand{\bits}{\mbox{bits}}
\newcommand{\GB}{\mbox{GB}}

\newcommand{\muop}{\mbox{$\mu$op}}
\newcommand{\cycle}{\mbox{cy}}
\newcommand{\cycles}{\mbox{cy}}
\newcommand{\eos}{~.}
\newcommand{\cma}{~,}

\begin{document}


\title{Exploring performance and power properties of modern multicore chips via 
   simple machine models}

\author{G. Hager, J. Treibig, J. Habich, and G. Wellein\\
Erlangen Regional Computing Center (RRZE)\\
Martensstr. 1, 91058 Erlangen, Germany}
\date{August 13, 2012}



\maketitle

\begin{abstract}
  Modern multicore chips show complex behavior with respect to
  performance and power. Starting with the Intel Sandy Bridge
  processor, it has become possible to directly measure the power
  dissipation of a CPU chip and correlate this data with the
  performance properties of the running code. Going beyond a simple
  bottleneck analysis, we employ the recently published
  Execution-Cache-Memory (ECM) model to describe the single- and
  multi-core performance of streaming kernels. The model refines the
  well-known roof\/line model, since it can predict the scaling and
  the saturation behavior of bandwidth-limited loop kernels on a
  multicore chip.  The saturation point is especially relevant for
  considerations of energy consumption. From power dissipation
  measurements of benchmark programs with vastly different
  requirements to the hardware, we derive a simple, phenomenological
  power model for the Sandy Bridge processor. Together with the ECM
  model, we are able to explain many peculiarities in the performance
  and power behavior of multicore processors, and derive guidelines
  for energy-efficient execution of parallel programs. Finally,
  we show that the ECM and power models can be successfully used
  to describe the scaling and power behavior of a lattice-Boltzmann 
  flow solver code.
\end{abstract}

\section{Introduction and related work}

The transition to multicore technology in mainstream scientific
computing has led to a plethora of new performance effects and
optimization approaches. Since Moore's law is alive and well, we may
expect growing core counts at least in the mid-term future, together
with the ubiquitous memory bandwidth bottleneck. Hence, sensitivity
(or lack thereof) to limited memory bandwidth is one of the essential
traits that can be used to characterize the performance of scientific
simulation codes. Any serious ``white-box'' performance modeling
effort with a claim to explain some particular performance aspect on
the core, chip, and node levels must be able to address the
interaction of code with inherent architectural bottlenecks, of which
bandwidth or, more generally, data transfer, is the most important
one. The balance metric \cite{schoenauer00} and its refined successor,
the roof\/line model~\cite{williams:roofline}, are very successful
instruments for predicting the bandwidth-boundedness of computational
loops. Beyond clear bottlenecks, successful performance modeling can
become very complex; even if cycle-accurate simulators were available,
understanding the cause of a particular performance issue may require
a grasp of computer architecture that the average computational
scientist, as a mere user of compute resources, does not have.  Simple
machine models that build on published, readily comprehensible information about
processor architecture can thus be very helpful also for the 
non-expert. Whenever the model fails to describe some performance
feature within some margin of accuracy, there is the opportunity to 
learn something new about the interaction of hardware and software.
The Execution-Cache-Memory (ECM) model~\cite{th09} is a starting
point for refined performance modeling of streaming loop kernels
that also allows a well-founded prediction of the saturation point
in case of strong bandwidth limitation, which was previously addressed
in a more phenomenological way~\cite{Suleman:2008:FTP:1353534.1346317}.
Performance modeling is certainly not limited to the single node, and
there are many examples of successful large-scale modeling efforts
\cite{Hoisie:2000,Nudd:2000,Kerbyson:sc01,pop2005}. 
However, the chip is where the ``useful code'' that solves an actual
problem is executed, and this is where insight into
performance issues starts. 

This paper is restricted to the multicore chip level. Using the Intel
Sandy Bridge processor as the target platform, we apply the ECM model
to simple streaming loops and to a well-optimized lattice-Boltzmann
flow solver. We show that the model is able to describe how the
single-thread performance and the intra-chip scaling come about, and
to predict the number cores needed to saturate the memory
interface. Such knowledge has substantial impact on the productive use
of hardware, because any parallelism that does not lead to a
measurable performance improvement must be regarded as overhead.

Besides performance aspects, considerations on power dissipation of
multicore chips have become more popular in recent years; after all,
concerns about the power envelope of high-end processors have stimulated
the multicore transition in the first place.  It is highly
conceivable that some future processor designs will not be faster but
``only'' more power-efficient than their predecessors. This
development has already started in the mobile and embedded market, but
will soon hit scientific computing as well. The latest x86
architectures have many power efficiency features built in, such as
the ability to go to ``deep sleep'' states, to quickly switch off
parts of the logic that are currently unused, and to speed up or slow
down the clock frequency together with the core voltage. A lot of
research has been devoted to using these features creatively, e.g.,
with dynamic voltage/frequency scaling (DVFS) or dynamic concurrency
throttling (DCT)
\cite{Horvath:2008:MEM:1454115.1454153,Suleman:2008:FTP:1353534.1346317,10.1109/TPDS.2012.95}, in order
to either save energy or limit the power envelope of highly parallel
systems. The Intel Sandy Bridge processor exposes some of its power
characteristics to the programmer by means of its ``Running Average
Power Limit'' (RAPL) feature. Among other things, it enables a
measurement of the power dissipation of the whole chip with
millisecond resolution and decent 
accuracy~\cite{10.1109/MM.2012.12,rapl-short}.

Starting from a thorough analysis of three benchmark codes, which have
strongly differing requirements on the hardware resources, we derive
simplified assumptions about the power dissipation behavior of serial
and parallel code on the Sandy Bridge architecture. These assumptions
are then used to construct a qualitative model for the ``energy to
solution'' metric, i.e., the energy it takes to solve a given
problem. It is qualitative in the sense that it accounts for the relevant 
effects without the claim for absolute accuracy.
The model describes the behavior of energy to solution with
respect to the number of active cores, the chip's clock frequency, the
serial, and -- if applicable -- the saturated performance levels.
Consequently, we are able to answer questions such as ``What is the
optimal number of cores for minimum energy to solution?'', ``Is it
more energy-efficient to use more cores at lower clock speed than
fewer cores at higher clock speed?'', ``When exactly does the `race to
idle' rule apply?'', ``Is it necessary to sacrifice performance in favor
of low energy consumption?'', and ``What is the influence of single-core
optimization on energy efficiency?''. Since the ECM model is able to
predict the scaling behavior of bandwidth-bound loops, there is a
natural connection between the performance and power models presented
in this work. Using the lattice-Boltzmann flow solver as an example,
we validate the power model and derive useful guidelines for the 
energy-efficient execution of this and similar codes.
Taken together, the performance and power dissipation properties of
modern multicore chips are an interesting field for modeling
approaches. We try to explore both dimensions here by using 
simple machine models that can be employed by domain scientists
but are still able to make predictions with sufficient accuracy to be
truly useful. 

This work is organized as follows: We first present the hardware and
software used for benchmarking in Sect.~\ref{sec:env}.  Then we
introduce the ECM model in Sect.~\ref{sec:ecm}, and apply it to
streaming kernels and to multicore-parallel execution.  In
Sect.~\ref{sec:power}, we study the power dissipation characteristics
of several benchmark codes and derive simplified assumptions, which
are subsequently used in Sect.~\ref{sec:powermodel} to construct an
elementary power model that makes predictions about the ``energy to
solution'' metric and its behavior with clock frequency, serial code
performance, and the number of cores used. We validate the model
against measurements on the chosen benchmark codes. A
lattice-Boltzmann flow solver is then used in Sect.~\ref{sec:lbm} as a
test case for both the ECM performance model and the power
model. Sect.~\ref{sec:conclusion} summarizes the paper and gives an
outlook to future research.

\section{Test bed and tools}\label{sec:env}

\subsection{Hardware}\label{sec:cpus}

Unless specified otherwise, we use a dual-socket eight-core Intel
Sandy Bridge EP platform for single-node measurements and validation
(Xeon E5-2680, 2.7\,\GHZ\ base clock speed, turbo mode up to
3.5\,\GHZ).  
The Intel Sandy Bridge microarchitecture 
contains
numerous enhancements in comparison to its predecessor, Westmere. The 
following features are most important for our analysis \cite{intelopt}:
\begin{itemize}
\item Compared to SSE, the Advanced Vector Extensions (AVX) instruction set extension
  doubles the SIMD register width from 128 to 256~bits. At the
  same time, the load throughput of the L1 cache is doubled from 
  16~\bytes\ to 32~\bytes\ per cycle, so that a Sandy Bridge core
  can sustain one full-width AVX load and one half-width AVX store
  per cycle. 
\item The core can execute one ADD and one MULT instruction per cycle
  (pipelined). With double-precision AVX instructions, this leads to a
  peak performance of eight \flops\ per cycle (sixteen at single
  precision). In general, the core has a maximum instruction
  throughput of six \muop{}s per cycle.  
\item The L2 cache sustains refills and evicts to and from L1
  at 256~bits per cycle (half-duplex). A full 64-\byte\ cache line 
  refill or evict thus takes two cycles. This is the same as on
  earlier Intel designs.
\item The L3 cache is segmented, with one segment per core. All
  segments are connected by a ring bus. Each segment has the 
  same bandwidth capabilities as the L2 cache, i.e., it can sustain 
  256~bits per cycle (half-duplex) for refills and evicts from L2.
  This means that the L3 cache is usually not a bandwidth bottleneck
  and streaming loop kernels show good scaling behavior when the 
  data resides in L3.
\item All parts of the chip, including the L3 cache (which is part
  of the ``Uncore''), run at the same clock frequency, which can
  be set to a fixed value in the range from 1.2--2.7\,\GHZ. The speed
  of the DRAM chips is constant and independent of the core clock.
\item One Xeon E5-2680 socket of our test systems has four DDR3-1333
  memory channels for a theoretical peak bandwidth of 42.7\,\GBS. In
  practice, 36\,\GBS\ can be achieved with the STREAM benchmark. This
  translates to 107~bits per core cycle (at base clock speed).
\item All power dissipation measurements include the ``package''
  metric only, i.e., they ignore the installed RAM. Preliminary
  results for the power dissipation of installed DIMMs are between 2
  and 9\,\W\ per socket (16\,\GB\ RAM in 4 DIMMs of 4\,\GB\ each),
  depending on the workload (memory-bound vs. cache-bound).
\end{itemize}

\subsection{Tools}\label{sec:tools}

We have used the Intel compiler Version 12.1 update 9 for 
compiling source codes.
Hardware counter measurements were performed with the
\lperfctr\ tool from the LIKWID tool suite
\cite{likwid-psti,likwidweb}, which, in its latest development 
release, can access the power information (via the RAPL interface
\cite{10.1109/MM.2012.12}) and the ``uncore'' events (i.e., L3 cache
and memory/QuickPath interface) on Sandy Bridge processors.

The LIKWID suite also contains \lbench\
\cite{likwid-bench-tools11}, a microbenchmarking framework that makes it easy to build and
run assembly language loop kernels from scratch, without the
uncertainties of compiler code generation.  \lbench\ was used
to validate the results for some of the streaming microbenchmarks 
in this work.

\section{A refined machine model for streaming loop kernels on multicore}\label{sec:ecm}

The majority of numerical codes in computational science are based on
streaming loop kernels, which show good data parallelism and are
largely compatible with the hardware prefetching capabilities of
modern multicore processors. For large data sets, such kernels are
often (but not always) bound by memory bandwidth, which leads to a
peculiar scaling behavior across the cores of a multicore chip: Up to
some critical number of cores $t_\mathrm s$ scalability is good, but
for $t>t_\mathrm s$ performance saturates and is capped by some
maximum level. Beyond the saturation point, the roof\/line model
\cite{williams:roofline} can often be used to predict the performance,
or at least its qualitative behavior with respect to problem
parameters, but it does not encompass effects that occur between the
cache levels. The ``Execution-Cache-Memory'' (ECM) model
\cite{th09,rabbitct} adds basic knowledge about the cache bandwidths
and organization on the multicore chip to arrive at a more accurate
description on the single-core level. Although the model can be used
to predict the serial and parallel performance of codes on multicore
processors, its main purpose is to develop a deeper understanding of
the interaction of code with the hardware. This happens when the model
\emph{fails to coincide} with the measurement (see
Sect.~\ref{sec:ecmserial} below).

In the following we give a brief 
account of this model, show how it connects to the roof\/line model,
apply it to parallel streaming kernels, and refine it to account for
some unknown (or undisclosed) properties of the cache hierarchy.
In Sect.~\ref{sec:lbm}, we apply the model to a lattice-Boltzmann
flow solver.

\subsection{The Execution-Cache-Memory (ECM) model: Single core}\label{sec:ecmserial}

The main premise of the ECM model is that the runtime of a loop is
composed of two contributions: (i) The time it takes to execute all
instructions, with all operands of loads or stores coming from or
going to the L1 data cache.  We call this ``core time.''  (ii) The
time it takes to transfer the required cache lines into and out of the
L1 cache. We call this ``data delays.'' The model further assumes that
hardware or software prefetching mechanisms are in place that hide all
cache transfer latencies. In this work, we additionally assume that
the cache hierarchy is strictly inclusive, i.e., that the lines in
each cache level are also contained in the levels below it. The model
can accommodate inclusive caches as well; see~\cite{th09} for examples.

Since all data transfers between cache levels occur in packets of one
cache line, the model always considers one cache line's worth of
work. For instance, if a double precision array must be read with unit
stride for processing, the basic unit of work in the model is eight
iterations at a cache line size of 64~\bytes.
The execution time for one unit of work is then composed of the
in-core part $T_\mathrm{core}$ and the data delays $T_\mathrm{data}$,
with potential overlap between them.  

$T_\mathrm{data}$ will reflect the time it takes to transfer data
to or from L1 through the memory hierarchy. This value will be larger
if the required cache line(s) are ``far away.'' 
Note that, since we have assumed perfect prefetching, this
is not a simple latency effect: It comes about because of limited
bandwidth and several possibly non-overlapping contributions. For
example, on a Sandy Bridge core, the transfer of a 64-\byte\ cache
line from L3 through L2 to L1 takes a maximum of four and a minimum of
two cycles (32-\byte\ wide buses between the cache levels), depending
on whether the transfers can overlap or not.  Furthermore, the L1 cache of
Intel processors is ``single-ported'' in the sense that, in any clock
cycle, it can either reload/evict cache lines from/to L2 or
communicate with the registers, but not both at the same time.

The core time $T_\mathrm{core}$ is more complex to estimate. In the
simplest case, execution is dominated by a clear bottleneck, such as
load/store throughput or pipeline stalls. Some knowledge about the
core microarchitecture, like the kind and number of execution ports or
the maximum instruction throughput, is helpful for getting a first
estimate.  For example, in a code that is completely dominated by
independent ADD instructions, the core time is, to first order,
determined by the ADD port throughput (one ADD instruction per cycle
on modern Intel CPUs). In a complex loop body, however, it is often
hard to find the critical execution path that determines the number of
cycles.  The Intel Architecture Code Analyzer (IACA) \cite{iaca} is a
tool that can derive more accurate predictions by taking dependencies
into account. See, e.g., \cite{rabbitct} for a detailed analysis of a
complex loop body with IACA.

Putting together a prediction for the overall execution time requires
making best- and worst-case assumptions about possible overlaps of the
different contributions described above. If the measured performance
data is far off those predictions, the model misses an important
architectural or execution detail, and must be refined.  A simple
example is the write-allocate transfer on a store miss: A naive model
for the execution of a store-dominated streaming kernel (like, e.g.,
array initialization \verb.A(:)=0.) with data in the L2 cache will
predict a bandwidth level that is much higher than 
the measurement. Only when taking into account that every cache line 
must be transferred to L1 first will the prediction be correct.


On many of today's multicore chips a single core cannot saturate the
memory interface, although a simple comparison of peak performance
vs.\ memory bandwidth suggests otherwise: An Intel Sandy Bridge core,
for example, has a (double precision) arithmetic peak performance of
$P_\mathrm{peak}=21.6\,\GFS$ at 2.7\,\GHZ, and the maximum memory
bandwidth of the chip is 36\,\GBS\ (see Sect.~\ref{sec:cpus}). This
leads to a machine balance of $B_\mathrm m=1.7\,\BF$.  The code
balance of the triad loop from the STREAM benchmarks
(\verb.A(:)=s*B(:)+C(:).) is $B_\mathrm c=16\,\BF$ (including the
write-allocate for \verb.A(:).), so the roof\/line model predicts a
strong limitation from memory bandwidth at a loop performance of
$P_\mathrm{peak}B_\mathrm m/B_\mathrm c\approx 2.3\,\GFS$. However,
the single-threaded triad benchmark only achieves about 940\,\MFS,
which corresponds to a bandwidth of 15\,\GBS\ (Fortran version,
compiler options
\verb+-O3 -openmp -xAVX -opt-streaming-stores never -nolib-inline+).

The ECM model attributes this fact to non-overlapping contributions
from core execution and data transfers: While loads and stores to the
three arrays are accessing the L1 cache, no refills or evicts between
L1 and L2 can occur. The same may be true for the lower cache levels
and even memory, so that memory bandwidth is not the sole
performance-limiting factor anymore. Core execution and transfers
between higher cache levels are not completely hidden and the maximum
memory bandwidth cannot be hit. However, when multiple cores access
main memory (or a lower cache level with a bandwidth bottleneck, like
the L3 cache of the Intel Westmere processor), the associated core times and data 
delays can overlap among the cores, and a point will be reached where
the bottleneck becomes relevant. Thus, it is possible to predict
when performance saturation sets in.

\subsection{The ECM model: Multicore scaling}\label{sec:ecmparallel}

The single-core ECM model predicts lower and upper limits for the
bandwidth pressure on all memory hierarchy levels. When the bandwidth
capacity of one level is exhausted, performance starts to
saturate~\cite{Suleman:2008:FTP:1353534.1346317}. On the Intel Sandy
Bridge processor, where the only shared bandwidth resource is the main
memory interface, this happens when the bandwidth pressure exceeds the
practical limit as measured, e.g., by a suitable multi-threaded
STREAM-like benchmark. We call this the ``saturation point.'' At this
point, the performance prediction from a balance model (see above)
works well.

The maximum main memory bandwidth is an input parameter for the model.
In principle it is possible to use the known parameters of the memory
interface and the DIMM configuration, but this is over-optimistic in
practice. For current Intel processors, the memory
bandwidth achievable with standard streaming benchmarks like the
McCalpin STREAM \cite{McCalpin1995,stream} is between 65 and 85\% of
the theoretical maximum. Architectural peculiarities, however, may then
impede optimal use of the memory interface with certain types of code.
One example are data streaming loops with a very large number of
concurrent load/store streams, which appear, e.g., in implementations
of the lattice-Boltzmann algorithm (see Sect.~\ref{sec:lbm} below).
The full memory bandwidth as seen with the STREAM benchmarks cannot be
achieved under such conditions. The reasons for this failure are 
unknown to us and will be further investigated.

\subsection{Validation via streaming benchmarks}\label{sec:ecmval}

We validate the ECM model by using the Sch\"onauer vector
triad~\cite{schoenauer00} as a throughput benchmark (see
Listing~\ref{lst:triadcode}) on the Sandy Bridge architecture. The
Sch\"onauer triad is similar to the STREAM triad but has one
additional load stream.
\begin{lstlisting}[caption={Pseudo-code for the vector triad throughput benchmark, including performance measurement. The actual benchmark loop is highlighted.},label=lst:triadcode,float=tp]
  double precision, dimension(:),allocatable :: A,B,C,D
! Intel-specific: 512-byte alignment of allocatables
!DEC$ ATTRIBUTES ALIGN: 512 :: A,B,C,D%\label{l:aa}%
  
  call get_walltime(S)
 
!$OMP PARALLEL PRIVATE(A,B,C,D,i,j)
...
     do j=1,R
! Intel-specific: Assume aligned moves
!DEC$ vector aligned%\label{l:va}%
!DEC$ vector temporal%\label{l:vt}%
        %do i=1,N\label{l:vts}%
           %A(i) = B(i) + C(i) * D(i)%
        %enddo\label{l:vte}%
        ! prevent loop interchange
        if(A(N/2).lt.0) call dummy(A,B,C,D)
     enddo
!$OMP END PARALLEL
  
  call get_walltime(E)
  
  WT = E-S
\end{lstlisting} 
Note that there is no real work sharing in the benchmark loop (lines
\ref{l:vts}--\ref{l:vte}), since the purpose of the code is to fathom
the bottlenecks of the architecture. The code is equipped with Intel
compiler directives to point out some crucial choices: All array
accesses are aligned to suitable address boundaries (lines \ref{l:aa}
and \ref{l:va}) to allow for aligned MOV instructions, which are
faster on some architectures. Furthermore, the generation of
nontemporal store instructions (``streaming stores'') is prevented
(line \ref{l:vt}).

\subsubsection{Single-core analysis}

\begin{figure}[tbp] \centering
\includegraphics*[height=0.32\textheight]{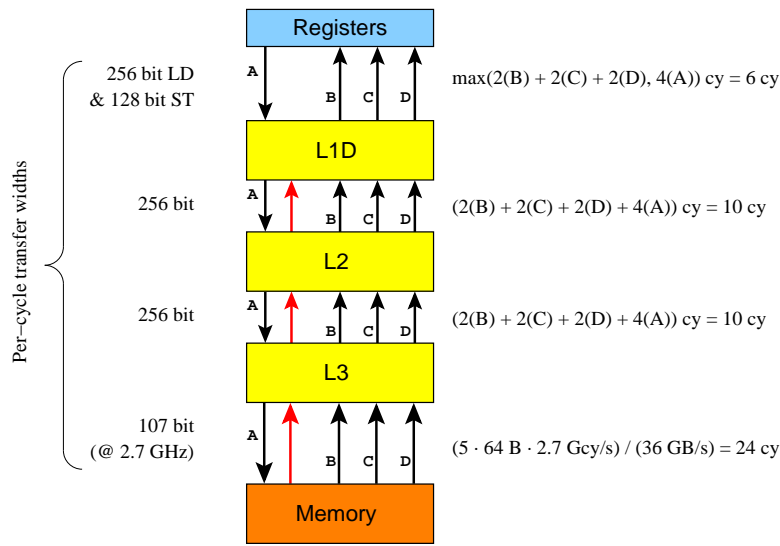}
\caption{\label{fig:triad_model}Single-core ECM model for the
Sch\"onauer triad benchmark (\texttt{A(:)=B(:)+C(:)*D(:)}) on Sandy
Bridge. The indicated cycle counts refer to eight loop iterations,
i.e., a full cache line length per stream. The transfer width per
cycle for refills from memory to L3 is derived from the measured
STREAM bandwidth limit of 36\,\GBS.  Red arrows indicate
write-allocate transfers.}
\end{figure}
All loop iterations are independent. Six full-width AVX
loads and two full-width AVX stores are required to execute one unit
of work (eight scalar iterations, or sixteen \flops, respectively).
From the microarchitectural properties described in
Sect.~\ref{sec:cpus} we know that this takes six cycles, since the
stores can be overlapped with the loads. In Fig.~\ref{fig:triad_model}
this is denoted by the arrows between the L1D cache and the registers.
The floating-point instructions do not constitute a bottleneck,
because only two ADDs and two MULTs are needed. 

Hence, the code has an overall instruction throughput of two
\muop{}s per cycle (four being the limit). The in-core
performance is limited by load/store throughput, and we have
$T_\mathrm{core}=6\,\cycles$.  Neglecting the loop counter and branch
``mechanics'' is justified because its impact can be minimized by
inner loop unrolling.

Due to the write misses on array \verb.A(:)., an additional cache line
load has to be taken into account whenever the data does not fit into
the L1 cache. This is indicated by the red arrows in
Fig.~\ref{fig:triad_model}. Ten cycles each are needed for the data
transfers between L2 and L1, and between L3 and L2, respectively.
As mentioned above, the memory bandwidth limit of 36\,\GBS\
leads to a per-cycle effective transfer width of 107\,\bits, which
adds another 24\,\cycles\ to $T_\mathrm{data}$.

\begin{figure}
\centering
\includegraphics*[height=0.34\textheight]{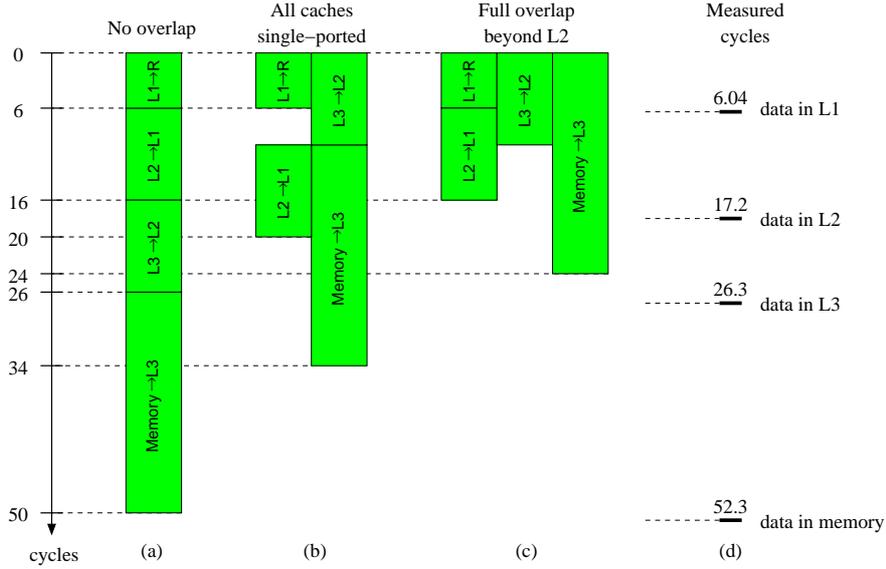}
\caption{\label{fig:triad_ovl} (a)--(c): Single-core timeline visualizations
  of the ECM model with cycle estimates for eight iterations (length of one 
  cache line) of the 
  Sch\"onauer triad benchmark on Sandy Bridge, with different overlap assumptions:
  (a) no overlap between all contributions from data transfers, (b) overlap
  under the condition that all caches are single-ported, i.e., can only
  talk to one immediately neighboring cache level at a time, (c) full overlap
  of all cache line transfers beyond L2.  
  For comparison the rightmost column (d) shows measurements in cycles per 
  eight iterations at the base clock frequency 
  of 2.7\,\GHZ, with the working set residing in different memory
  hierarchy levels.}
\end{figure}
Figure \ref{fig:triad_ovl} shows how the different parts can
be put together to arrive at an estimate for the execution time.
In the worst case, the contributions to $T_\mathrm{data}$ can neither
overlap with each other nor with $T_\mathrm{core}$, leading to 
$T=50\,\cycles$ for data in memory, $26\,\cycles$ for L3,
and 16\,\cycles\ for L2 (see Fig.~\ref{fig:triad_ovl}a). 
On the other hand, assuming full overlap beyond the L2 cache
(see Fig.~\ref{fig:triad_ovl}c), the minimum possible
execution times are $T=24\,\cycles$, 16\,\cycles, and 16\,\cycles,
respectively. The only well-known fact in terms of overlap
is that the L1 cache is single-ported, which is why no overlap
is assumed even in the latter case. 

Assuming the non-overlap condition for all cache levels,
we arrive at the situation depicted in Fig.~\ref{fig:triad_ovl}b:
Contributions can only overlap if they involve a mutually exclusive
set of caches. We then arrive at a prediction of $T=34\,\cycles$
for in-memory data, 20\,\cycles\ for L3, and again 16\,\cycles\
for L2 (the latter cannot be shown in the figure). 

Figure \ref{fig:triad_ovl}d shows measured execution times for
comparison. We must conclude that there is no overlap taking
place between any contributions to $T_\mathrm{core}$ and 
$T_\mathrm{data}$. Note that this analysis is valid for a
single type of processor, and that other microarchitectures 
may show different behavior. It must also be stressed that
the existence of overlap also depends strongly on the type
of code (see also the next section).

\subsubsection{Multicore scaling}\label{sec:mc_triad}

All resources in the Sandy Bridge processor chip, except for the
memory interface, scale with the number of cores. Hence we predict
good scalability of the benchmark loop up to eight cores, if the data
resides in the L3 cache. Indeed we see a speedup of 7.4 from
one to eight cores. In the memory-bound regime we expect
scalability up to the bandwidth limit of 107\,\bits/\cycle, which is
a factor of 2.09 larger than the single-core bandwidth prediction
of $320\,\bytes/50\,\cycles=51.2\,\bits/\cycle$.
The performance of the Sch\"onauer triad loop should thus saturate
at three cores, with a small speedup from two to three. 
\begin{figure}
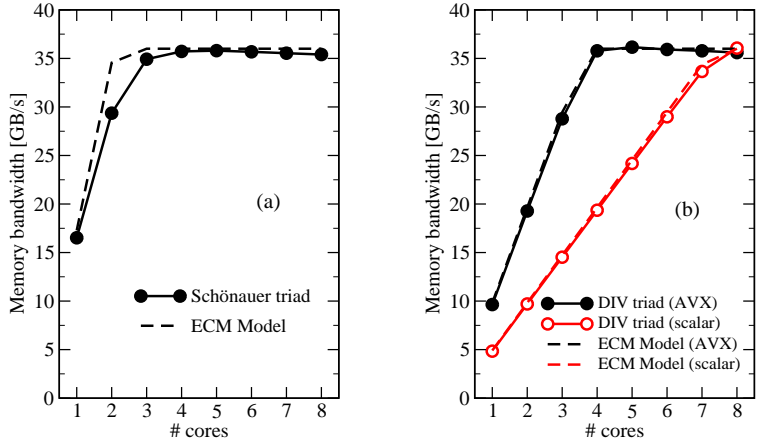

\centering
\includegraphics*[height=0.26\textheight]{sb-triad-scaling}\hspace{1cm}
\includegraphics*[height=0.26\textheight]{sb-dtriad-scaling}
\caption{\label{fig:sb-triad-scaling}Multicore
  scaling of (a) the memory-bound Sch\"onauer triad benchmark and 
  (b) the modified triad with a divide (\texttt{A(:)=B(:)+C(:)/D(:)}), 
  in comparison with the 
  corresponding ECM models (dashed lines) on a 2.7\,\GHZ\ 
  Sandy Bridge chip. The model for (a) assumes no overlap, 
  while the model in (b) assumes full overlap of $T_\mathrm{core}$
  with $T_\mathrm{data}$.}
\end{figure}
Figure \ref{fig:sb-triad-scaling} shows a comparison of the 
model with measurements on a Sandy Bridge chip at 2.7\,\GHZ.
The model tracks the overall scaling behavior well, especially
the number of cores where saturation sets in. Note that we 
expect the same
general characteristics for all loop kernels that are strongly
load/store-bound in the L1 cache if the data traffic volume
between all cache levels is roughly constant. At $t=2$, the model 
over-predicts the performance by about 15\%. This deviation
is yet to be investigated.

For comparison, we modified the vector triad code so that a divide is
executed instead of a multiplication between arrays \verb.C(:).  and
\verb.D(:).. The throughput of the double-precision full-width AVX
divide on the Sandy Bridge microarchitecture is 44 cycles if no
shortcuts can be taken by the hardware \cite{intelopt}, while the
throughput of a scalar divide is 22 cycles. All required loads and
stores in the L1 cache can certainly be overlapped with the
large-latency divides, leading to an in-core execution time of
88\,\cycles\ and 172\,\cycles, respectively, for the AVX and scalar
variants. In this case, the single-portedness of the L1 cache is not
applicable, since the in-core code is not load/store-bound. Even
if no overlap takes place in the rest of the hierarchy, the $10+10+24=44$ 
additional cycles for data transfers (see Fig.~\ref{fig:triad_model}) 
can be hidden behind the 
in-core time. The results in Fig.~\ref{fig:sb-triad-scaling}b show a 
very good agreement of the ECM model with the measurements.

\section{Power dissipation and performance on multicore}\label{sec:power}

In this section we will first investigate the power dissipation
properties of the Sandy Bridge processor by studying several benchmark
codes. We then develop a simple power model and derive from it the
most interesting features for the ``energy to solution'' metric with
respect to clock frequency, number of cores utilized, and serial code
performance.

\subsection{Power and performance of benchmarks vs. active cores}

A couple of test codes were chosen, each of which shows a somewhat
typical behavior for a certain class of applications. We report
performance, CPI (cycles per instruction), and power dissipation on a
Sandy Bridge EP (Xeon E5-2680) chip, with respect to the number of 
cores used. The ``turbo mode'' feature was deliberately ignored.

\begin{figure}
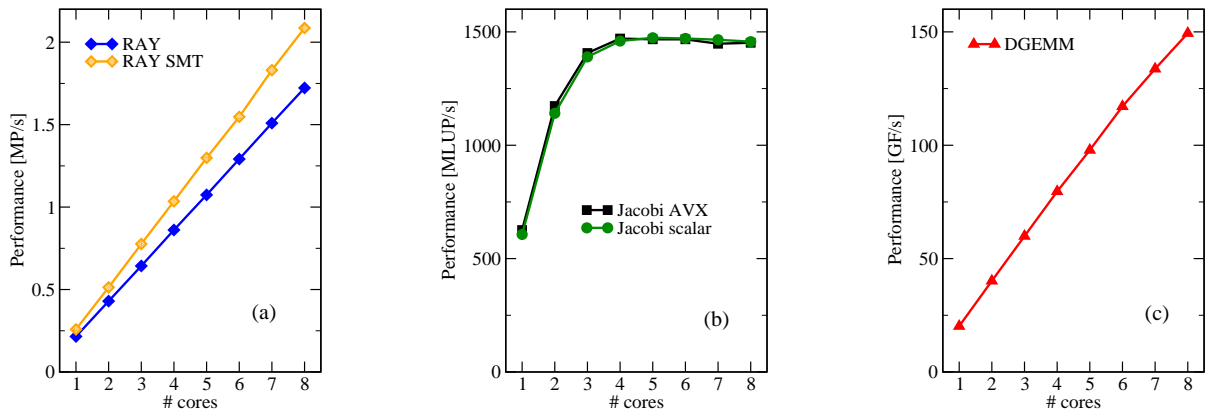

\centering
\includegraphics*[height=0.24\textheight]{ray_perf_vs_cores}\hfill
\includegraphics*[height=0.24\textheight]{j2d_perf_vs_cores}\hfill
\includegraphics*[height=0.24\textheight]{dgemm_perf_vs_cores}
\caption{\label{fig:perf_vs_cores}Performance of the benchmark codes
  on a Sandy Bridge chip with respect to the number of active cores at
  the base frequency of 2.7\,\GHZ.}
\end{figure}
\begin{figure}
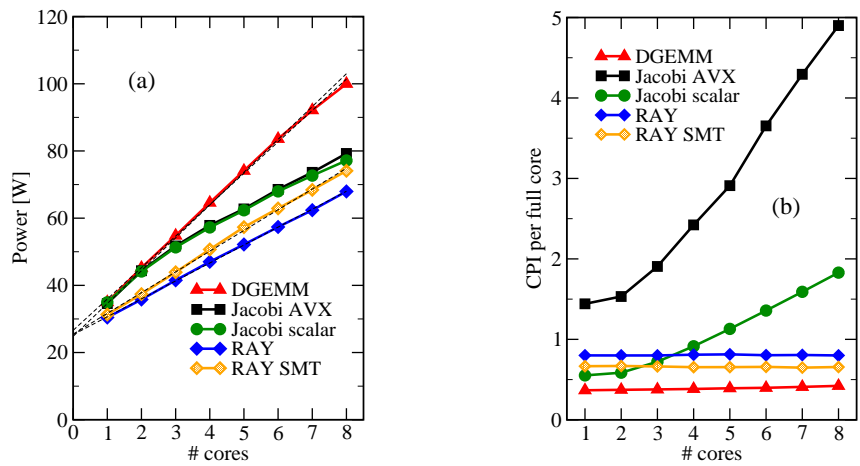

\centering
\includegraphics*[height=0.27\textheight]{power_vs_cores}\hspace{2cm}
\includegraphics*[height=0.27\textheight]{cpi_vs_cores}
\caption{\label{fig:w_cpi_vs_cores}(a) Power dissipation and (b) cycles per
  instruction of the benchmark codes with respect to the number of
  cores used, at the base frequency of 2.7\,\GHZ.}
\end{figure}
In the following, we briefly describe the benchmark codes together
with performance and power data with respect to the number of cores used
(see Figs.~\ref{fig:perf_vs_cores} and \ref{fig:w_cpi_vs_cores}).

\subsubsection{RAY} 
is a small, MPI-parallel, master-worker style ray-tracing program,
which computes an image of $15000^2$ gray-scale pixels of a scene
containing several reflective spheres. Performance is reported in
million pixels per second (\MPS).  

Scalability across the cores of a multicore chip is perfect (see
Fig~\ref{fig:perf_vs_cores}a), since all data comes from L1 cache, load
imbalance is prevented by dynamic work distribution, and there is no
synchronization apart from infrequent communication of computed tiles
with the master process, which is pinned to another socket and thus
not taken into account in the analysis below. The code is purely
scalar and shows a mediocre utilization of the core resources with a
CPI value of about $0.8$ (see Fig~\ref{fig:w_cpi_vs_cores}b). It
benefits to some extent from the use of simultaneous multi-threading
(SMT), which reduces the CPI to $0.65$ per (full) core for a speedup
of roughly 15\%.  At the same time, power dissipation grows by about
8\% and is roughly linear in the number of cores used for both cases
(see Fig.~\ref{fig:w_cpi_vs_cores}a).

\subsubsection{Jacobi}
is an OpenMP-parallel 2D Jacobi smoother (four-point stencil) used
with an out-of-cache data set ($4000^2$ lattice sites at double
precision).  Being load/store-dominated with an effective code balance
of 6\,\BF\ \cite{Hager10}, it shows the typical saturation behavior
described by the ECM model for streaming codes. Performance is
reported in million lattice site updates per second (\MLUPS), where
one update comprises four \flops. Hence, we expect a saturation
performance of 6\,\GFS\ or 1500\,\MLUPS\ on a full Sandy Bridge chip,
which is fully in line with the measurement (see
Fig.~\ref{fig:perf_vs_cores}b).  

This benchmark was built in two
variants, an AVX-vectorized version and a scalar version, to see the
influence of data-parallel instructions on power dissipation. Both
versions have very similar scaling characteristics, with the scalar
code being slightly slower below the saturation point, as expected.
The performance saturation is also reflected in the CPI rate
(Fig.~\ref{fig:w_cpi_vs_cores}b), which shows a linear slope after
saturation. Surprisingly, although there is a factor of 2.5 in terms
of CPI between the scalar and AVX versions, the power dissipation
hardly changes (Fig.~\ref{fig:w_cpi_vs_cores}a).  Beyond the
saturation point, the slope of the power dissipation decreases
slightly, indicating that a large CPI value is correlated with lower
power. However, the relation is by no means inversely proportional,
just as for the RAY benchmark.

\subsubsection{DGEMM} 
performs a number of multiplications between two dense double
precision matrices of size $5600^2$, using the thread-parallel Intel
MKL library that comes with the Intel compiler (version 10.3 update
9). Performance is reported in \GFS. 

The code scales almost perfectly with a speedup of 7.5 on eight cores,
and reaches about 86\% of the arithmetic peak performance on the full
Sandy Bridge chip at a CPI of about 0.4 (i.e, 2.5 instructions per
cycle).  Power dissipation is almost linear in the number of cores
used (Fig.~\ref{fig:w_cpi_vs_cores}a).

DGEMM achieves the highest power dissipation of all codes considered
here.  Note that at the base frequency of 2.7\,\GHZ, the thermal
design power (TDP) of the chip of 130\,\W\ is not nearly reached, not
even with the DGEMM code. With turbo mode enabled (3.1\,\GHZ\ at eight
cores) we have measured a maximum sustained power of 122\,\W. The
Sandy Bridge chip can exceed the TDP limit for short time
periods~\cite{10.1109/MM.2012.12}, but this was not investigated here.

Surprisingly, the power dissipation of DGEMM is identical to the
Jacobi code (scalar and AVX versions) as long as the latter is not
bandwidth-bound, whereas the RAY benchmark draws about 15\% less power
at low core counts. We attribute this to the mediocre utilization of
the execution units in RAY, where some long-latency floating-point
divides incur pipeline stalls, and the strong utilization of the full
cache hierarchy by the Jacobi smoother.

\subsection{Power and performance vs. clock frequency for all benchmarks}

\begin{figure}
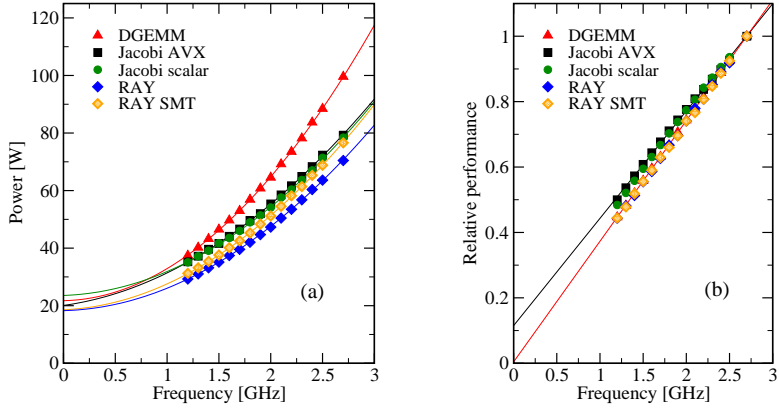

\centering
\includegraphics*[height=0.24\textheight]{power_sep1}\hspace{1cm}
\includegraphics*[height=0.24\textheight]{perf_1T_vs_f}
\caption{\label{fig:power_perf_sep1}(a) Power dissipation of a Sandy Bridge
  chip with respect to clock speed for the benchmark codes. All
  eight physical cores were used in all cases, and all 16 virtual cores
  for the ``RAY SMT'' benchmark. The solid lines are least-squares 
  fits to a second-degree polynomial in $f$. (b) Relative performance 
  versus clock speed with respect to the $2.7\,\GHZ$ level of single-core
  execution for the benchmark codes. Two processes
  on one physical core were used in case of RAY SMT. The solid lines are 
  linear fits to the Jacobi AVX and DGEMM data, respectively.}
\end{figure}

Figure \ref{fig:power_perf_sep1}a shows the power dissipation of all codes
with respect to the clock frequency ($f=1.2\ldots2.7\,\GHZ$) when all
cores are used (all virtual cores in case of the SMT variant of RAY).
The Sandy Bridge chip only allows for a ``global'' frequency setting,
i.e., all cores run at the same clock speed.
The solid lines are least-squares fits to a second-degree polynomial,
\bq\label{eq:powerf}
W(f) = W_0 + w_1f + w_2f^2\cma
\eq
for which the coefficient of the linear term is very small compared
to the constant and the quadratic term. The quality of the fit 
suggests that the dependence of dynamic power dissipation on frequency
is predominantly quadratic with $7\,\W/\GHZ^2<w_2<10\,\W/\GHZ^2$, depending on
code characteristics. Note that one would naively expect a cubic
dependence in $f$, if the core voltage is adjusted to always
reflect the lowest possible setting at a given frequency. Since
we cannot look into the precise algorithm that the hardware uses
to set the core voltage, we use the observed quadratic function as 
phenomenological input to the power model below, without questioning 
its exact origins. The conclusions we draw from the model
would not change qualitatively if $W(f)$ were a cubic polynomial, or
any other monotonically increasing function with a positive second
derivative.

The ``base power'' $W_0\approx 23\,\W$
(leakage) is largely independent of the type of code, which 
is plausible; however, one should note that an extrapolation to 
$f=0$ is problematic here, so that the estimate for $W_0$ is very
rough.

In Fig.~\ref{fig:power_perf_sep1}b, we show the single-core performance
of all benchmarks with respect to clock frequency, normalized to the
level at $f=2.7\,\GHZ$. As expected, the codes with near-perfect
scaling behavior across cores show a strict proportionality of
performance and clock speed, since all required resources run with the
core frequency. In case of the Jacobi benchmark the linear
extrapolation to $f=0$ has a small $y$-intercept, because resources are
involved that are clocked independently of the CPU cores. The ECM
model predicts this behavior if one can assume that the maximum
bandwidth of the memory interface is constant with varying frequency.

\begin{figure}
\centering
\includegraphics*[width=0.5\textwidth]{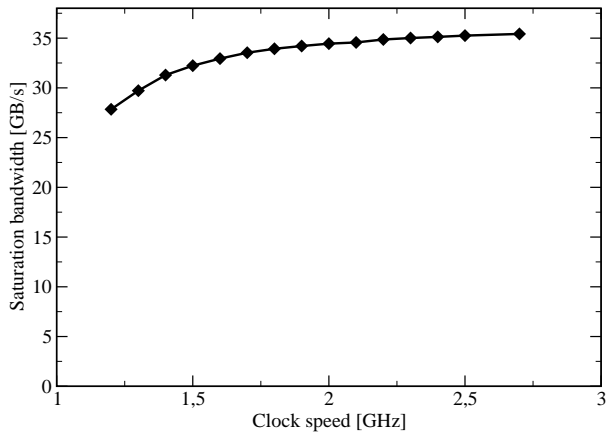}
\caption{\label{fig:membw_vs_clock_sb}Maximum memory bandwidth (saturated)
  versus clock frequency of a Sandy Bridge chip.}
\end{figure}

Figure \ref{fig:membw_vs_clock_sb} shows the saturated memory
bandwidth of a Sandy Bridge chip with respect to clock speed. 
If we assume that the core frequency should not influence the memory
interface, we have no explanation for the drop in bandwidth below
about $1.7\,\GHZ$: The ECM model predicts constant bandwidth for a
streaming kernel like, e.g., the Sch\"onauer triad (one may speculate
whether a slow Uncore clock speed could cause a lack of outstanding
requests to the memory queue, reducing achievable bandwidth).
For the purpose
of developing a multicore power model, we neglect these effects and
assume a strictly linear behavior (with zero $y$-intercept) of
performance vs.\ clock speed in the non-saturated case.

\subsection{Conclusions from the benchmark data}\label{sec:benchconc}

In order to arrive at a qualitative model that connects the power and
performance features of the multicore chip, we will now draw some 
generalizing conclusions from the data that was discussed in the previous 
sections.

From Fig.~\ref{fig:power_perf_sep1}a, we conclude that the dynamic
power dissipation is a quadratic polynomial in the clock frequency and
parameterized by $w_2$ in (\ref{eq:powerf}). $w_2$ depends strongly on
the type of code executed, and there is some (inverse) correlation
with the CPI value (see Fig.~\ref{fig:w_cpi_vs_cores}), but a simple 
mathematical relation cannot be derived. The linear part $w_1$ is 
generally small compared to $w_2$. 

A linear extrapolation of power dissipation vs.\ the number of active
cores to zero cores (dashed lines in Fig.~\ref{fig:w_cpi_vs_cores}a)
shows that the baseline power of the chip is $W_0\approx25\,\W$,
independent of the type of running code. In case of the
bandwidth-limited Jacobi benchmark only the one- and two-core data
points were considered in the extrapolation. The result for $W_0$ is
also in line with the quadratic extrapolations to zero clock frequency
in Fig.~\ref{fig:power_perf_sep1}a. Note that $W_0$, as a 
phenomenological model parameter, is different from
the documented ``idle power'' of the chip, which is considerably
lower due to advanced power gating mechanisms. 

From the same data we infer a linear dependence of power dissipation 
on the number of active cores $t$ in the non-saturated regime,
\bq
W(f,t)  = W_0 + (W_1f + W_2f^2)t\cma
\eq
so that $w_{1,2}=t\cdot W_{1,2}$.  Although the power per core rises
more slowly in the saturation regime, we regard this as a second-order
effect and neglect it in the following: The fact that a core is active
has much more influence on power dissipation than the characteristics
of the running code.

As Fig.~\ref{fig:w_cpi_vs_cores}a indicates, using both hardware threads 
(virtual cores) on a physical Sandy Bridge
core increases power dissipation due to the improved utilization of
the pipelines. The corresponding performance increase depends on the
code, of course, so it may be more power-efficient to ignore the SMT
threads.  In case of the RAY code, however, the increase in power is
over-compensated by a larger boost in performance, as shown in 
Fig.~\ref{fig:perf_vs_cores}a.  See Sect.~\ref{sec:etsbench} for 
further discussion.

One of the conclusions from the ECM model was that, in the
non-saturated case, performance is proportional to the core's clock
speed.  Fig.~\ref{fig:power_perf_sep1}b suggests that this
true for the scalable benchmarks, and approximately true
also for saturating codes like Jacobi.

\section{A qualitative power model}\label{sec:powermodel}

Using the measurements and conclusions from the previous section we
can now derive a simple power model that describes the overall power
properties of a multicore chip with respect to number of cores used,
the scaling properties of the code under consideration, and the clock
frequency. As a starting point we choose the ``energy to solution''
metric, which quantifies the energy required to solve a certain
compute problem. Hence, we restrict ourselves to strong scaling
scenarios. This is not a severe limitation, since weak scaling is
usually applied in the massively (distributed-memory) parallel case,
where the relevant scaling unit is a node or a ccNUMA domain (which is
usually a chip).  The optimal choice of resources and execution
parameters on the chip level, where the pertinent bottlenecks are
different, are done at a constant problem size.

The following basic assumptions go into the model:
\begin{enumerate}
\item The whole $N_\mathrm{c}$-core chip consumes some baseline power
  $W_\mathrm 0$ when powered on, which is independent of the number of 
  active cores and of the clock speed.
\item An active core consumes a dynamic power of $W_1f+W_2f^2$. 
  We will also consider deviations from
  some baseline clock frequency $f_0$ such that $f=(1+\Delta \nu)f_0$, with
  $\Delta\nu=\Delta f/f_0$.
\item At the baseline clock frequency, the serial code under consideration 
  runs at some performance $P_0$. As long as there is no bottleneck,
  the performance is linear in the number of cores used, $t$, and the 
  normalized clock speed, $1+\Delta\nu$. The latter dependence 
  will not be exactly linear if some part of the hardware (e.g., the 
  outer-level cache) runs at its own clock speed.
  In presence of a bottleneck (like, e.g., memory
  bandwidth), the overall performance with respect to $t$ is
  capped by some maximum value $P_\mathrm{max}$:
\bq
  P(t)=\min\left((1+\Delta \nu)tP_0,P_\mathrm{max}\right)\eos
\eq
  Here we assume that performance scales with clock frequency until 
  $P_\mathrm{max}$ is reached.
\end{enumerate}
Since time to solution is inverse performance, the energy to solution becomes
\bq\label{eq:etosol}
E=\frac{W_0+(W_1f+W_2f^2)t}{\min\left((1+\Delta\nu)tP_0,P_\mathrm{max}\right)}\eos
\eq

\subsection{Minimum energy with respect to number of cores used}\label{sec:ets_cores}

Due to the assumed saturation of performance with $t$, we have to 
distinguish two cases:

\paragraph{Case 1: $(1+\Delta\nu)tP_0<P_\mathrm{max}$} 

Performance is linear in the number of cores, so that (\ref{eq:etosol}) becomes
\bq\label{eq:e_below}
E=\frac{W_0+(W_1f+W_2f^2)t}{(1+\Delta\nu)tP_0}\cma
\eq
and $E$ is a decreasing function of $t$:
\bq
\frac{\partial E}{\partial t}=-\frac{W_0}{(1+\Delta\nu)t^2P_0}<0\eos
\eq
Hence, the more cores are used, the shorter the execution time and the smaller 
the energy to solution.

\paragraph{Case 2: $(1+\Delta\nu)tP_0>P_\mathrm{max}$}

Performance is constant in the number of cores, hence
\bq\label{eq:e_above}
E=\frac{1}{P_\mathrm{max}}\left(W_0+(W_1f+W_2f^2)t\right)
\eq
\bq
\Rightarrow\quad
	\frac{\partial E}{\partial t}=
	\frac{1}{P_\mathrm{max}}\left(W_1f+W_2f^2\right)>0\eos
\eq
In this case, energy to solution grows with $t$, with a slope that is
proportional to the dynamic power, while the time to solution stays
constant; using more cores is thus a waste of energy.

For codes that show performance saturation at some $t_\mathrm s$, it follows
that energy (and time) to solution is minimal just at this point:
\bq\label{eq:tsat}
t_\mathrm s=\frac{P_\mathrm{max}}{(1+\Delta\nu) P_0}\eos
\eq
If the code scales to the available number of cores, case 1 applies 
and one should use them all.

\subsection{Minimum energy with respect to serial code performance}\label{sec:e_vs_p}

Since the serial code performance $P_0$ only appears in the
denominator of (\ref{eq:etosol}), increasing $P_0$ leads to decreasing
energy to solution unless $P=P_\mathrm{max}$. A typical example for
this scenario is the SIMD vectorization of a bandwidth-bound code:
Using data-parallel instructions (such as SSE or AVX) will
generally reduce the execution overhead in the core, so that $P_0$
grows and the saturation point $P_\mathrm{max}$ is reached at smaller
$t$ (see (\ref{eq:tsat})). Consequently, the potential for saving
energy is twofold: When operating below the saturation point, 
optimized code requires less energy to solution. At the saturation
point, one can get away with fewer active cores to solve the problem 
at maximum performance.

\subsection{Minimum energy with respect to clock frequency}\label{sec:ets_clock}

We again have to distinguish two cases:

\paragraph{Case 1: $(1+\Delta\nu)tP_0<P_\mathrm{max}$} 

Energy to solution is the same as in (\ref{eq:e_below}) and $f=(1+\Delta\nu)f_0$, so that
\bq
\frac{\partial E}{\partial\Delta\nu}=
\frac{f_0^2}{tP_0}\left(W_2 t-\frac{W_0}{f^2}\right)\eos
\eq
The derivative is positive for large $f$; setting it to zero 
and solving for $f$ thus yields the frequency for
minimal energy to solution:
\bq\label{eq:fopt}
f_\mathrm{opt} = \sqrt{\frac{W_0}{W_2t}}\eos
\eq
A large baseline power $W_0$ forces a large clock frequency 
to ``get it over with'' (``clock race to idle''). Depending on
$W_0$ and $W_2$, $f_\mathrm{opt}$ may be larger than the
highest possible clock speed of the chip, so that there
is no energy minimum. This may be the case if one includes 
the rest of the system  in the analysis (i.e., memory, disks, 
etc.). On the other hand, a large dynamic
power $W_2$ allows for smaller $f_\mathrm{opt}$, since the loss in performance
is over-compensated by the reduction in power dissipation.
The fact that $f_\mathrm{opt}$ does not depend on $W_1$ just reflects our
assumption that the serial performance is linear in $f$.

Since $t$ appears in the denominator in (\ref{eq:fopt}), 
it is tempting to conclude that
a clock frequency reduction can be compensated by using more cores,
but the influence on $E$ has to be checked by inserting
$f_\mathrm{opt}$ from (\ref{eq:fopt}) into (\ref{eq:e_below}): 
\bq\label{eq:eminatfopt}
E(f_\mathrm{opt})=\frac{f_0}{P_0}\left(2\sqrt{\frac{W_0W_2}{t}}+W_1\right)
\eq
This confirms our conjecture that, below the saturation point, more 
cores at lower frequency save energy. At the same time, performance
at $f_\mathrm{opt}$ is
\bq\label{eq:patfopt}
P(f_\mathrm{opt})=\frac{f_\mathrm{opt}}{f_0}tP_0 = 
  \frac{P_0}{f_0}\sqrt{\frac{W_0t}{W_2}}\cma
\eq
hence it grows with the number of cores: trading cores for
clock slowdown does not compromise time to solution.

However, if $t$ is fixed, (\ref{eq:patfopt}) also tells us that, if
$f_\mathrm{opt}<f_0$, performance will be smaller than at the base
frequency $f_0$, although the energy to solution is also smaller.
This may be problematic if $t$ cannot be made larger to
compensate for the loss in performance. In this case the
energy to solution metric is insufficient and one has to choose
a more appropriate cost function, such as energy multiplied by
runtime:
\bq\label{eq:costRT}
C=\frac EP = \frac{W_0+(W_1f+W_2f^2)t}{\left((1+\Delta\nu)tP_0\right)^2}\eos
\eq
Differentiating $C$ with respect to $\Delta\nu$
gives
\bq
\frac{\partial C}{\partial\Delta\nu}=
-\frac{2W_0+W_1ft}{(f/f_0)^3P_0^2}<0\cma
\eq
because $W_0,\,W_1>0$. 
Hence, a higher clock speed is always better if $C$ is chosen
as the relevant cost function. Note that a large baseline power
$W_0$ emphasizes this effect, e.g., when the whole system
is taken into account (see also above in the discussion of
$f_\mathrm{opt}$).

\paragraph{Case 2: $(1+\Delta\nu)tP_0>P_\mathrm{max}$} 

Beyond the saturation point, energy to solution is the same as in 
(\ref{eq:e_above}), so it grows with the frequency: The clock should be as
slow as possible. Together with the findings from case 1 this means that
minimal energy to solution is achieved when using all available cores,
at a clock frequency which is so low (if possible) that the saturation
point is right at $t=N_\mathrm{c}$.

These results reflect the popular ``clock race to idle'' rule, which
basically states that a processor should run at maximum frequency to
``get it over with'' and go to sleep as early as possible to
eventually save energy. Using the energy to solution behavior as
derived above, we now know how this strategy depends on the number of
cores used and the ratio of baseline and dynamic power. ``Clock race
to idle'' makes sense only in the sub-saturation regime, and when
$f<f_\mathrm{opt}$. Beyond $f_\mathrm{opt}$ (if such frequencies are
allowed), the quadratic dependence of power on clock speed will waste
energy. Beyond the saturation point, i.e., if $t>t_\mathrm s$, lower
frequency is always better.

\subsection{Validation of the power model for the benchmarks}\label{sec:etsbench}

The multicore power model has been derived from the benchmarks' power
dissipation using considerable simplifications. 
Hence, we will now check whether the
conclusions drawn above are still valid
for the benchmark codes when looking at the measured energy
to solution data with respect to the number of active cores,
the clock speed, and the single-core performance.

\begin{figure}
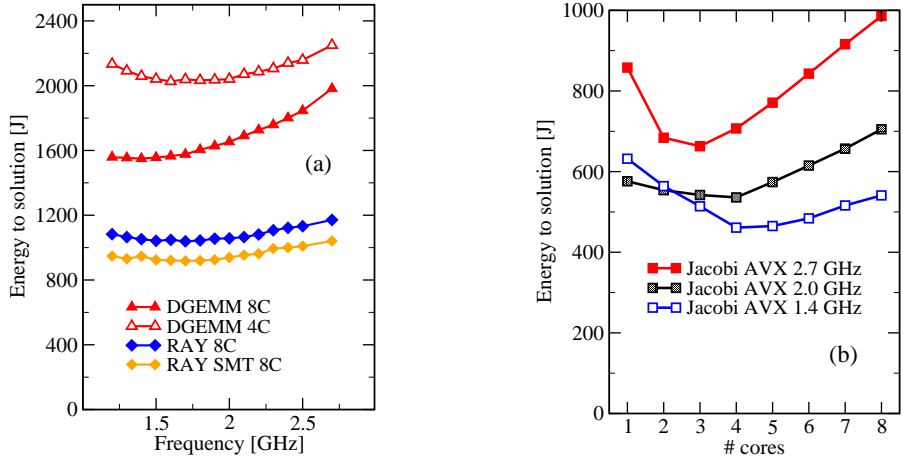

\centering
\includegraphics*[height=0.27\textheight]{ets_dgemm_ray_vs_f}\hspace{2cm}
\includegraphics*[height=0.27\textheight]{j2d_ets_vs_cores}
\caption{\label{fig:ets}Energy to solution for
  (a) the scalable benchmarks DGEMM (eight and four cores) and RAY 
     (eight cores) versus clock frequency
  on a Sandy Bridge socket and 
  (b) the Jacobi AVX benchmark versus number of cores at
  different core frequencies. }
\end{figure}

Figure~\ref{fig:ets} shows energy to solution measurements for 
the scalable codes (Fig.~\ref{fig:ets}a) and the Jacobi AVX 
benchmark (Fig.~\ref{fig:ets}b) versus clock frequency and
number of cores, respectively. 

Comparing the frequency for minimum energy to solution between DGEMM
and RAY at eight cores (solid symbols in Fig.~\ref{fig:ets}a),
we can identify the behavior predicted by (\ref{eq:fopt}): A
large dynamic power factor $W_2$ leads to lower $f_\mathrm{opt}$. The
SMT version of RAY consumes more power than the standard version, but,
as anticipated above, the larger performance leads to lower energy to
solution: Better resource utilization on the core, i.e., optimized
code, saves energy; this provides another possible attitude towards the
``race to idle'' rule. Given the huge amount of optimization potential
that is still hidden in many production codes on highly
parallel systems, we regard this view as even more relevant than
optimizing clock speed for a few percent of energy savings.

Eq.~(\ref{eq:fopt}) predicts a larger optimal frequency $f_\mathrm{opt}$
at fewer cores, which is clearly visible
when comparing the four- and eight-core energy data for DGEMM in
Fig.~\ref{fig:ets}a (solid vs.\ open triangles). 
At the same time, fewer cores also lead to
larger minimum energy to solution at $f_\mathrm{opt}$, which was 
shown in (\ref{eq:eminatfopt}).

The Jacobi benchmark shows all the expected features of a code whose
performance saturates at a certain number of cores $t_\mathrm s$: As
predicted by the ECM model, the saturation point is shifted to a
larger number of cores as the clock frequency goes down; we have
derived in Sect.~\ref{sec:ets_cores} that this is the point at which
energy to solution is minimal. Lowering the frequency, $t_\mathrm s$
gets larger, but energy to solution decreases (see
(\ref{eq:eminatfopt})). When $t>t_\mathrm s$, more cores and higher
clock speed both are a waste of energy. At $t<t_\mathrm s$ the Jacobi
code is largely frequency-bound and there is an optimal frequency
$f_\mathrm{opt}\approx 2\,\GHZ$ with minimal energy to solution. Here we substantiate
the prediction from Sect.~\ref{sec:ets_clock} that ``clock race to
idle'' is largely counterproductive if we look at the chip's
power dissipation only. See also
Sect.~\ref{sec:lbm_perfmodel} for a discussion of ``race to idle''
in the context of a lattice-Boltzmann CFD solver.

In conclusion, although considerable simplifications have been 
made in constructing the model (\ref{eq:etosol}), it is able 
to describe the qualitative
behavior of the benchmark applications with respect to energy 
to solution.

\section{Case study: A D3Q19 lattice-Boltzmann fluid solver}\label{sec:lbm}

\subsection{The lattice-Boltzmann algorithm and its implementation}

The widely used class of lattice Boltzmann models with BGK
approximation of the collision process \cite{lba:wolf-gladrow:2000,lba:succi:2001b,lba:chen:1998,lba:qian:1992} is based on the evolution
equation
\begin{equation}
  f_i(\vec{x}+\vec{e}_i\delta t,\,t+\delta t) = f_i(\vec{x},\,t) 
       -\frac{1}{\tau}\left[f_i(\vec{x},\,t) -
                            f_i^\mathrm{eq}(\rho, \vec{u}) \right] 
                          \qquad i=0 \ldots N.\quad
\label{eq:Boltzmann}
\end{equation}
Here, $f_i$
denotes the particle distribution function (pdf), which represents the
fraction of particles located in timestep $t$ at position $\vec{x}$
and moving with the microscopic velocity $\vec{e}_i$. The relaxation
time $\tau$ determines the rate of approach to local equilibrium
and is related to the kinematic viscosity of the fluid.  
The equilibrium state $f_i^\mathrm{eq}$ itself is a low Mach number
approximation of the Maxwell-Boltzmann equilibrium distribution
function. It depends only on the macroscopic values of the fluid
density~$\rho$ and the flow velocity~$\vec{u}$.
%
%
Both can be easily obtained as the first moments of the particle 
distribution function.
The discrete velocity vectors $\vec{e}_i$ arise from the $N$ chosen
collocation points of the velocity-discrete Boltzmann equation and
determine the basic structure of the numerical grid. A typical
3D discretization is the D3Q19 model~\cite{lba:qian:1992}, which
uses 19 discrete velocities (collocation points). 



Each timestep $(t\rightarrow t+\delta t)$ consists of the following
steps, which are repeated for all cells of the computational domain:
\begin{itemize}
\item Compute the local macroscopic flow quantities $\rho$ and
  $\vec{u}$ from the distribution functions, $\rho = \sum_{i=0}^N f_i$
  and $\vec{u} = \frac{1}{\rho}\sum_{i=0}^N f_i \vec{e}_i$.
\item Calculate the equilibrium distribution $f_i^\mathrm{eq}$
  from the macroscopic flow quantities (see~\cite{lba:qian:1992} for
  the equation and parameters) and execute the ``collision''
  (relaxation) process, 
    $f_i^* (\vec{x},\,t^*) = f_i(\vec{x},\,t) 
       -\frac{1}{\tau}\left[f_i(\vec{x},\,t) -
                            f_i^\mathrm{eq}(\rho, \vec{u}) \right]$,
  where the superscript ``*'' denotes the post-collision state (``collide step'').
\item Propagate the $i=0 \ldots N$ post-collision states
  $f_i^* (\vec{x},\,t^*)$ to the appropriate neighboring cells
  according to the direction of $\vec{e}_i$, resulting in $f_i
  (\vec{x}+\vec{e}_i\delta t,\,t+\delta t)$, i.e., the values of the
  next timestep (``stream step'').
\end{itemize}
The first two steps are computationally intensive and involve values
of the local node only. They can be easily combined and will be
denoted ``collide step'' in the following. The third step is a
direction-dependent uniform shift of data in memory involving no
floating-point computation and will be denoted ``streaming step''.
A fourth step, the so-called
``bounce-back'' rule \cite{lba:ziegler:1993}, is incorporated as an
additional part of the stream step and ``reflects'' the
distribution functions at the interface between fluid and solid cells,
resulting in an approximate no-slip boundary condition at walls.
Since bounce-back is only required at the interfaces, it 
has a minor impact on runtime
if the ratio of solid to fluid cells is small. Hence, we will omit
it for the analysis below.

\begin{lstlisting}[caption={Pseudo-code for the SoA D3Q19 LBM ``split'' benchmark kernel, with split-off loops highlighted. The bounce-back rule is omitted.},label=lst:lbmcode,float=tp,firstnumber=1]
  double precision, dimension(:,:,:,:,:),allocatable :: PDF
  double precision, dimension(:),allocatable :: loc_dens
  double precision, dimension(:),allocatable :: ux,uy,uz
  ...
!$OMP PARALLEL SHARED(A)
!$OMP DO
...
do z=1,Nz
 do y=1,Ny
  %do x=1,Nx\label{l:inner_start}%
   ! Caclulate density
   loc_dens(x) =  PDF(x,y,z,C,t)
                 + PDF(x-1,y,z,E,t)
                 + PDF(x,y-1,z,N,t)
                 ...
                 + PDF(x,y+1,z+1,BS,t)
      
   !calculate moments and store to ux,uy,uz
   ux(x)= PDF(x-1,y,z,E,t)-PDF(x+1,y,z,W,t) 
        + PDF(x-1,y-1,z,NE,t) +PDF(x+1,y-1,z,NW,t)...
     
   uy(x)= PDF(x,y-1,z,N,t)-PDF(x,y+1,z,S,t) 
        + PDF(x-1,y-1,z,NE,t) -PDF(x-1,y+1,z,SE,t)...
     
   uz(x)=...
  %enddo%
  
  ! 19 store loops: tN = new time, t = old time
  %do x=1,Nx\label{l:wback}%
   ! reuse ux,uy,uz and loc_dens
   ! Propagation to local node
   PDF(x,y,z,C,tN) =  PDF(x,y,z,C,t)   *... ux(x)...
  %enddo%
  
  %do x=1,Nx%
   ! reuse ux,uy,uz and loc_dens
   ! Propagation to local node
   PDF(x,y,z,E,tN) =  PDF(x-1,y,z,E,t) *... ux(x)...
  %enddo%

  ! 17 more propagation loops here ...
  ...
   %\label{l:inner_end}%
  enddo
enddo
!$OMP END DO
!$OMP END PARALLEL
  
\end{lstlisting} 

\subsubsection{Combined Collide-Stream kernel} 
The LBM algorithm builds on the two physical processes, \emph{collision} and \emph{streaming}.
For this report we use the so-called ``pull'' scheme, i.e., the
streaming step is followed by the collision.  
Using appropriate data storage schemes, e.g., two disjoint arrays,
it is therefore possible to do
both steps within a single loop. This reduces the overall data
transfer between main memory and CPU to one load, one store, and one write-allocate per
lattice cell update and distribution function.
%

\subsubsection{Data layout} 
Another basic decision is the choice of the data layout for the arrays
holding the particle distribution functions. Here we consider a full
matrix representation, where the pdf array covers the full simulation
domain independent of the character of the cells (fluid vs.\ solid).
Choosing rectangular 3D computational domains this requires a pdf
array with at least four dimensions (three spatial coordinates
$(x,y,z)$ and the discretization stencil ($Q=0,\ldots,18$)). We only
consider the structure-of-arrays (SoA) approach (\verb.PDF(x,y,z,Q)., using
Fortran notation), as it provides maximum attainable performance
without the need for spatial blocking at large problem 
sizes~\cite{hpc:wellein:2006,hpc:Zeiser:2009:ppl:paper}.

\subsubsection{Loop splitting}
The SoA implementation requires 19 load and 19 store streams, addressing
completely different parts of the pdf array concurrently.
The large number of competing store streams is problematic for the
underlying memory architecture due to, e.g., the limited number of store
buffers. Thus it is often beneficial to separate the store streams by
updating a smaller number of directions (1--5) at a time. This can
easily be implemented by having separate $x$ (inner) loops for the different
chunks of store streams and holding some intermediate results in small
buffer arrays (which comes at the cost of additional cache traffic and 
instructions). Listing~\ref{lst:lbmcode} shows the pseudo-code for 
a D3Q19 SoA split-store LBM kernel. 

\subsubsection{SIMD processing}
The optimizations of data layout and loop splitting allow for a large
fraction of the maximum performance (in double precision) according to
the roof\/line model. For single precision computation, however,
performance does not double as would be expected. This clearly
indicates the need to improve the use of processor/on-chip
resources by full vectorization. Considering the predictions of the
ECM and power models with respect to single-threaded performance
and energy to solution, respectively, SIMD vectorization will
not only improve the single-precision performance but also,
more importantly, shift the saturation point to a smaller number
of cores, leading to lower energy to solution also in double
precision. We have thus implemented a vectorized version of the
LBM algorithm from Listing~\ref{lst:lbmcode} using AVX SIMD
intrinsics. 

\subsection{Performance results}\label{sec:lbmperf}

\begin{figure}[tbp]
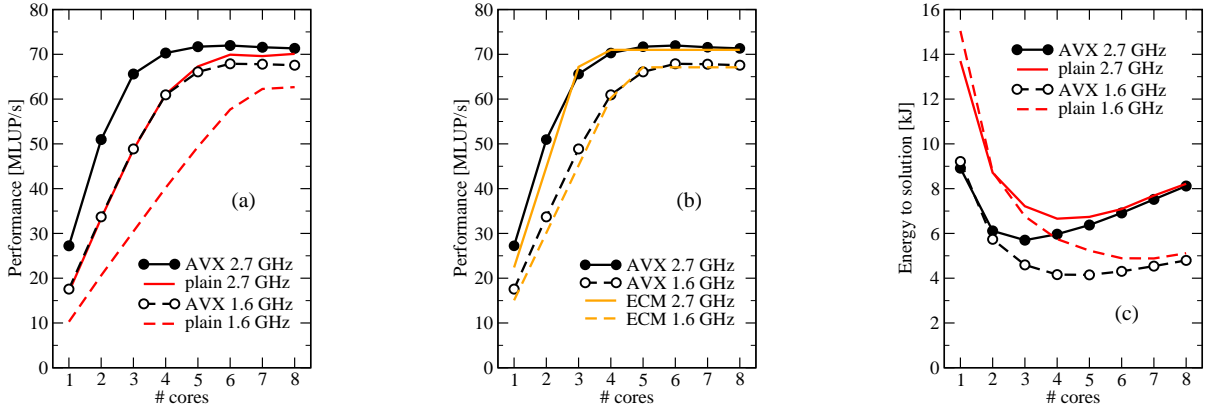

\centering
\includegraphics*[height=0.24\textheight]{lbm_perf_vs_cores}\hfill
\includegraphics*[height=0.24\textheight]{lbm_avx_vs_ecm}\hfill
\includegraphics*[height=0.24\textheight]{lbm_ets_vs_cores}
\caption{\label{fig:lbm_p_ets_vs_cores}(a) Performance 
   of the AVX and plain (scalar) variants of the LBM solver
   versus number of cores on a Sandy Bridge socket. (b) Comparison of the AVX
 LBM code with the non-overlap ECM model. (c) Energy to solution of the
 code variants from (a) versus number of cores.}
\end{figure}
We use a cubic domain of $228^3$ fluid cells (about $3.7\,\GB$ of memory)
without obstacles at double precision as a benchmark case for LBM.
Figure~\ref{fig:lbm_p_ets_vs_cores}a shows the performance of the
plain (scalar) and AVX implementations of the LBM algorithm with
respect to the number of cores at 2.7\,\GHZ\ and 1.6\,\GHZ,
respectively. As expected, at 2.7\,\GHZ\ the AVX version is faster below saturation
and also saturates earlier, but arrives at the same maximum
performance as the scalar code at larger core counts.  At reduced
clock speed the scalar code shows almost perfect scaling up to seven
cores. The memory bandwidth degradation observed with simple streaming
benchmarks (see Fig.~\ref{fig:membw_vs_clock_sb}) can also be seen here:
About 5--10\% of saturated performance is lost at 1.6\,\GHZ, depending
on the code variant.

\subsection{Application of the performance model}\label{sec:lbm_perfmodel}

The Intel Architecture Code Analyzer (IACA) \cite{iaca} can compute
pure execution time ($T_\mathrm{core}$) predictions from generated
assembly code, using some built-in knowledge about the
microarchitecture as input. It outputs several numbers: (i) The
minimum number of cycles during which each of the core's issue ports
is busy, assuming no dependencies between instructions. The largest of
those numbers (the ``total throughput'') is the minimum execution time
of the analyzed code in cycles. (ii) A refined prediction based on
dependencies between instructions along the critical execution path
(the ``performance latency''). We generally use the total throughput
in order to arrive at a prediction that is an absolute upper
performance limit.

\begin{lstlisting}[float,
caption={Intel Architecture Analyzer output for the AVX LBM kernel on Intel Sandy Bridge.},label=lst:Iaca-output-snb-avx-split,basicstyle=\tiny\ttfamily]
Analysis Report
---------------
%Total Throughput: 216 Cycles%;           Throughput Bottleneck: %Port2\verb._.DATA, Port3\verb._.DATA%
Total number of Uops bound to ports:  656
Data Dependency Latency:    81 Cycles;  Performance Latency:    271 Cycles

Port Binding in cycles:
----------------------------------------------------------------
|  Port  |  0  -  DV |  1  |  2  -  %D%  |  3  -  %D%  |  4  |  5  |
----------------------------------------------------------------
| Cycles |  98 |  0  | 114 | 141 | %216% | 180 | %216% |  50 |  97 |
----------------------------------------------------------------
\end{lstlisting}
\begin{lstlisting}[float,
caption={Intel Architecture Analyzer output for the scalar LBM kernel on Intel Sandy Bridge.},label=lst:Iaca-output-snb-plain-full-split,basicstyle=\tiny\ttfamily]
Analysis Report
---------------
%Total Throughput: 223 Cycles%;           Throughput Bottleneck: %Port1%
Total number of Uops bound to ports:  899
Data Dependency Latency:    122 Cycles; Performance Latency:    259 Cycles

Port Binding in cycles:
----------------------------------------------------------------
|  Port  |  0  -  DV |  %1%  |  2  -  D  |  3  -  D  |  4  |  5  |
----------------------------------------------------------------
| Cycles | 122 |  0  | %223% | 199 | 168 | 199 | 168 |  62 |  94 |
----------------------------------------------------------------
\end{lstlisting}
Listings \ref{lst:Iaca-output-snb-avx-split} and
\ref{lst:Iaca-output-snb-plain-full-split} show the IACA analysis for
the plain and AVX-based LBM kernel variants, respectively.  The tool
was given the complete assembly code for the statements
between lines \ref{l:inner_start} and \ref{l:inner_end} in 
Listing~\ref{lst:lbmcode}.
It predicts a minimum in-core execution time of 216\,\cycles\ per
AVX-vectorized LBM loop iteration, i.e., 432\,\cycles\ per unit of
work (eight lattice updates), the bottleneck being the LOAD ports. For
the scalar code, the minimum prediction is 223\,\cycles, i.e.,
1784\,\cycles\ per unit of work, with the ADD port as the bottleneck.
The fact that the AVX prediction is almost exactly four times faster
than the scalar one is pure coincidence, however: Inspection of the
assembly code revealed that in some cases the Intel compiler fails to
generate full-width AVX loads and stores from SIMD intrinsics but
chooses half-width (128\,\bit) variants instead, especially for the
write-back loops starting at line \ref{l:wback} in
Listing~\ref{lst:lbmcode}. This is also the reason why the AVX code is
limited by the LOAD port, whereas the scalar code is limited by the
ADD port.

The inner loops require 19 load and store streams to the outer memory
hierarchy. The problem size in $x$ direction was chosen such that the
loads and stores from and to the auxiliary arrays \verb.loc_dens(:).,
\verb.ux(:)., \verb.uy(:)., and \verb.uz(:). are handled within the L1
cache. From a data traffic point of view, the LBM thus resembles 19
parallel, contiguous array copy operations. As already mentioned in
Sect.~\ref{sec:ecmparallel}, under such conditions the memory
interface of the Sandy Bridge chip fails to achieve its maximum
bandwidth according to the STREAM benchmarks, but saturates at about
$32.3\,\GBS$ at $2.7\,\GHZ$ clock speed ($30.6\,\GBS$ at $1.6\,\GHZ$). 
We use these values as  upper limits in the following.

\begin{figure}[tbp] \centering
\includegraphics*[height=0.32\textheight]{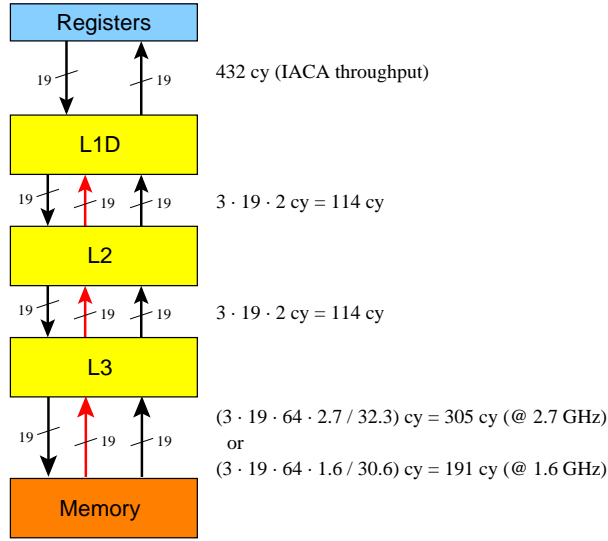}
\caption{\label{fig:lbm_model}Single-core ECM model for the
lattice-Boltzmann kernel with AVX on Sandy
Bridge at $2.7$ and $1.6\,\GHZ$. One arrow represents the transfer
of 19 cache lines for the 19 distribution functions.
Red arrows indicate write-allocate transfers. The two predictions for the 
memory transfers emerge from the limit given by a multi-stream
copy benchmark ($32.3$ and $30.6\,\GBS$, respectively).}
\end{figure}

Together with the IACA analysis we arrive at the single-core cycle
counts displayed in Fig.~\ref{fig:lbm_model} (we only discuss the more
interesting AVX variant here, since the scalar code is strongly
dominated by in-core execution).
Assuming that no overlap takes place within the hierarchy,
$965\,\cycles$ ($851\,\cycles$) are needed for one unit of work, which amounts to a
performance of $22.4\,\MLUPS$ ($15\,\MLUPS$) at $2.7\,\GHZ$ ($1.6\,\GHZ$). 
Figure~\ref{fig:lbm_p_ets_vs_cores}b shows a
comparison of the model with measurements. At both frequencies the 
scaling behavior 
is tracked quite accurately, including the saturation point,
but obviously the no-overlap assumption is slightly too pessimistic
for this code and there must be some (small) overlap,
most probably between execution and data transfers.

We will see in the next section that the accurate prediction of the
saturation point has significance for the power behavior.

\subsection{Application of the power model}

Since LBM is bandwidth-bound much along the same lines as the
Jacobi solver (albeit with a lower code balance), we expect
lowest energy to solution at the performance saturation point.
The strong difference in scalability between the AVX and scalar
variants also allows a direct assessment of the impact on 
single-thread code optimization on energy consumption. 

Figure~\ref{fig:lbm_p_ets_vs_cores}c shows the energy to solution for
a fixed number of LBM updates versus the number of cores used. As 
expected, the general behavior is very similar to the Jacobi benchmark (see
Fig.~\ref{fig:ets}), and exactly as predicted by the power model 
for bandwidth-bound codes:
Minimum energy to solution is achieved at the saturation point $t=t_\mathrm s$
predicted by (\ref{eq:tsat}) (e.g., at three to four cores for AVX at 2.7\,\GHZ\,
and at about five cores at 1.6\,\GHZ). However, using more cores at lower 
frequency leads to a 5--10\,\% loss of saturated performance; a refined 
cost model similar to (\ref{eq:costRT}) should be used in this
case to determine whether the loss in performance is outweighed 
by the gain in energy efficiency.

If there is no clear saturation
(as with the plain variant at 1.6\,\GHZ), the available number of
cores is barely sufficient to reach an energy minimum, which
is expected from the power model for scalable codes, (\ref{eq:e_below}). 
In the serial case $t=1$, 
where the chip's power consumption is dominated by
the baseline power $W_0$, a slow clock is counterproductive
and leads to higher energy to solution (dashed lines above solid lines),
as was predicted in Sect.~\ref{sec:ets_clock}. 
This is the point where ``clock race to idle'' makes sense
for this code, but using less than $t_\mathrm s$  cores
would be extremely wasteful since 
energy to solution is more than twice as large as
at the saturation point.

Finally, comparing the AVX and plain versions of the LBM
implementation at the saturation point, we recover the prediction from
Sect.~\ref{sec:e_vs_p} that optimizing code saves energy: ``Racing to
idle'' with better code is always beneficial.

\section{Conclusion}\label{sec:conclusion}

\subsection*{Summary}
In this paper, we have applied the Execution-Cache-Memory (ECM) model to 
simple streaming kernels and to a lattice-Boltzmann flow solver.
We have shown that this model can describe the scaling properties 
of bandwidth-bound codes on a multicore chip better than 
a simple bottleneck analysis.
Although our analysis on the Sandy Bridge EP processor suggests that
there is no overlap between data transfers on different levels of the
memory hierarchy, this may not be a general result and must be
investigated on other microarchitectures and for more loop kernels.

Starting from measured performance and power dissipation properties of
three benchmarks with different demands on the hardware, we have
furthermore established a simple model for ``energy to solution'' of
parallel codes on a multicore chip. The model also takes typical
saturation properties for bandwidth-bound loop kernels into account
and thus builds implicitly on the ECM performance model. The behavior
of energy to solution is qualitatively described by the model with
respect to the number of active cores, the performance of the
single-threaded code, and the clock frequency, despite substantial
simplifications from the underlying measurements: We have derived a
definite condition for the usefulness of ``clock race to idle,'' i.e.,
running with a high clock frequency to complete a computation as fast
as possible, which applies when the code scales well with the number
of cores. If there is a saturation point with respect to core count,
lowest energy to solution is achieved when the number of threads is
chosen so that saturation sets in, at the lowest possible clock speed.
Finally, better code performance always improves energy to solution,
and is thus the most important aspect in saving energy on
parallel systems.
These results have been confirmed using the simple benchmarks
on which the model was built, and by applying the model to
a lattice-Boltzmann flow solver. 

\subsection*{Outlook}
The specific issues of over-optimistic performance prediction
below the saturation point for strongly load/store-bound kernels, and the 
memory bandwidth degradation with loop kernels
that have a large number of concurrent load/store streams is still
poorly understood, and will be studied in more detail in the future.
The ECM model may also be extended to accommodate the dependence
of achievable memory bandwidth on the core clock frequency. 
Since the Sandy Bridge processor also allows for a measurement of the
power dissipation of connected RAM modules, this contribution
to the overall energy to solution metric must be incorporated into
the power model, although we do not expect substantial changes in the
conclusions. 

Future work will encompass the application of the performance and
power models, in order to fathom their domain of applicability, to
more processor architectures and application codes and to massively
parallel applications. We plan to extend the models to codes that
exhibit more complex performance patterns, such as load imbalance and
synchronization or communication overhead.  In the light of upcoming
architectures that are able to set the clock frequencies of individual
cores, the ECM and power models will have to be derived and validated
again under those more complex boundary conditions.

We also expect future processor and system designs to make power
information more accessible so that comprehensive power measurement
functions can be implemented in the \lperfctr\ tool and used in
analysis.

\section*{Acknowledgments}

 This work was supported by the German 
Federal Ministry for Education and Research (BMBF) under 
grants no. 01IH08003A (project SKALB) and 01IH11002A (project
hpcADD), and by the Competence
Network for Scientific High Performance Computing in 
Bavaria (KONWIHR) via the project OMI4papps.
We are indebted to Intel Germany for providing a
Sandy Bridge EP test platform through an early access
program. Useful discussions with 
Ian Steiner (Intel) are gratefully acknowledged.

\bibliographystyle{wileyj}
\bibliography{rrze,lbm}

\end{document}